\documentclass[twocolumn,prb,aps,preprintnumbers,amsmath,amssymb
,floatfix]{revtex4}

\usepackage{amsmath}
\usepackage{graphicx}
\usepackage{textcomp}
\usepackage{color}
\usepackage{amsfonts}
\usepackage{epsfig}
\usepackage{hyperref}

\newcommand{\arctanh}[1]{\operatorname{arctan}}

\begin{document}

\title{Multiscale Modeling of Nanowire-based Schottky-Barrier Field-Effect Transistors for Sensor Applications} 
\date{\today} 

\author{Daijiro Nozaki$^{1}$}
\email[Corresponding Author, Electronic address:]{ daijiro.nozaki@nano.tu-dresden.de}
\author{Jens Kunstmann$^{1}$}
\author{Felix Z\"{o}rgiebel$^{1}$}
\author{\\ Walter M. Weber$^{2}$}
\author{Thomas Mikolajick$^{2}$}
\author{Gianaurelio Cuniberti$^{1,3}$}
\affiliation{$^{1}$Institute for Materials Science and Max Bergmann Center of Biomaterials, 
TU Dresden, 01062 Dresden, Germany} 
\affiliation{$^{2}$NaMLab gGmbH. N\"{o}thnitzer Str.~64 01187 Dresden Germany}
\affiliation{$^{3}$Division of IT Convergence Engineering and National Center for
Nanomaterials Technology, POSTECH, Pohang 790-784, Republic of Korea}

\begin{abstract}
\noindent
We present a theoretical framework for the calculation of charge transport through nanowire-based Schottky-barrier field-effect transistors that is conceptually simple but still captures the relevant physical mechanisms of the transport process.  
Our approach combines two approaches on different length scales: 
(1) the finite elements method is used to model realistic device geometries and to calculate the electrostatic potential across the Schottky-barrier by solving the Poisson equation, and (2) the Landauer approach combined with the method of non-equilibrium Green's functions is employed to calculate the charge transport through the device.
%
%
Our model correctly reproduces typical I-V characteristics of field-effect transistors and the dependence of the saturated drain current on the gate field and the device geometry are in good agreement with experiments.
Our approach is suitable for one-dimensional Schottky-barrier field-effect transistors of arbitrary device geometry and it is intended to be a simulation platform for the development of nanowire-based sensors.
\end{abstract}


\maketitle

\section{\label{sec:intro}Introduction}
Over the last few decades one-dimensional (1D) semi-conducting silicon-nanowires (SiNWs) have been widely studied as potential building blocks for future electronic devices due to their excellent electrical performance, small size, and controllable bottom-up fabrication.\cite{Wagner,Westwater,CP2010,Rurali} 
Many types of SiNW devices such as ultrasensitive sensors,\cite{Cui,APA} photodetectors,\cite{photo} and bipolar field effect transistors (FETs)\cite{bipo,Koo} have been reported. The key issue for sensor applications is the down-scaling of FETs to one dimensional structures, such as nanowires. The attractive feature of nanowire-based FETs is that the binding of charged species can be directly monitored by the change in current through the NWs because of their high surface-to-volume ratios and small cross-sectional conduction pathways.

Recently Weber \textit{et al.} have reported dopant-free Schottky-barrier (SB) FETs consisting of intrinsic SiNWs working as a channel and $\mathrm{NiSi}_2$ nanowires working as source and drain contacts with gate lengths down to subphotolithographic values.\cite{weber,weber2,weber3,weber4} Their devices are based on single-crystalline SiNWs into which nickel atoms are diffusing from both ends of the wire, forming sharp interfaces between NiSi$_\mathrm{2}$ ``leads'' and a Si ``channel'' in the center. The channel length depends on the annealing time.
Measurements of their transport characteristics have shown the highest on-current and on-conductance values recorded to date for intrinsic SiNW-FETs.  It is advantageous that the silicon channel is dopant free since the noise caused by structural impurities is greatly reduced compared to doped nanowires. 
This feature is beneficial for the production of reliable SiNW-based biosensors.  

By modifying the surface of the SiNWs with DNA,\cite{Nano_Lett-04-51, Nano_Lett-04-245,Nano_Lett-08-1066,APA,Anal_Chem-79-3291} enzymes\cite{electrochem} or other chemical compounds\cite{Cui} many kinds of biosensors such as EnFET (enzyme FETs), Immuno-FETs,\cite{Biosens_Bioelec-accepted} and DNA-FETs have been demonstrated. 
Thus, SiNW-based SB-FETs are expected to provide a promising platform for biosensor applications. 

For a better understanding of the basic charge transport characteristics through the SiNW-based SB-FETs and the development of the NW-based biosensors, it is essential to systematically reveal the factors controlling the charge transport through a pristine FET device. 

Several approaches to describe the electronic transport through 1D SB-FETs have been proposed. 
Knoch \textit{et al.} developed quantum mechanical simulations of a ultrashort channel n-metail-oxide-semiconductor FETs (n-MOSFETs) on silicon on insulator (SOI)\cite{Knoch1}  and of a SB-FETs on SOI\cite{Knoch2} using real-space non-equilibrium Green's function formalism.
Heinze \textit{et al.} introduced a method to calculate the transport through carbon nanotube (CNT) SB-FETs using the Landauer formula and the WKB approximation.\cite{Heinze}
Pourfath \textit{et al.} extended this approach to cylindrical double-gate CNT SB-FETs.\cite{Pourfath}
Michetti \textit{et al.} expressed the conduction band by an analytical formula and calculated the transmission functions analytically.\cite{Michetti}
Appenzeller \textit{et al.} have investigated the transport properties of CNT-FETs and SiNW-FETs in detail.\cite{Appenzeller} They used a modified 1D Poisson equation and the non-equilibrium Green's function technique to calculate the  transmission function with the finite difference method in 1D.
Jimen\'{e}z \textit{et al.} expressed the conduction band profile by an analytical expression\cite{Jimenez07,Dubois} and calculated the transport properties of SB-FETs by using the WKB approximation, corrected with k-matching conditions for MSi2/Si(111) and MSi2/Si(100) (with M=Ni, Co, and Fe) interfaces.\cite{Dubois}

The majority of these approaches were applied to highly symmetric devices such as cylindrical FETs designed for integrated circuits. In these systems, the symmetry significantly reduces the computation time. However, to simulate realistic devices for sensor applications, a computationally cheap model that is applicable to arbitrary device geometries  is also desired. 
The multiscale model presented in this article was developed for this purpose. Our method combines two approaches on different length scales: (1) the finite elements method (FEM) is used to calculate the three-dimensional (3D) electrostatic potential across the Schottky-barrier by solving the Poisson equation for realistic device geometries. Then we extract the 1D potential profile along the axis of the NWs and use (2) the Landauer approach combined with the method of non-equilibrium Green's functions\cite{Landauer, Keldysh, Buttiker} to calculate the charge transport through the device. 
Our model is conceptually simple, computationally inexpensive, it uses only a few empirical parameters, and it is easily expandable. Since the method is non-atomistic the number of atoms in the system is irrelevant and therefore it is possible to simulate devices with dimensions ranging from a few nanometers up to some micrometers.

In this work we model pristine SB-FETs consisting of SiNWs and NiSi$_2$ NWs. We analyze the influence of the device geometries such as gate lengths, thickness of the insulators, and the gate voltages on the charge transport through the SB-FETs.
Despite the simplicity of the model, the numerical calculations show  I-V characteristics of typical conventional MOSFETs and the calculated saturation currents are in good agreement with the experimental results by Weber \textit{et al.}\cite{weber}

This paper is organized as follows. Section \ref{sec:theory} presents theoretical framework of multiscale modeling and section \ref{sec:methods} shows the computational details and parameter settings. In section \ref{sec:4}, we present dependence of the electrostatic potential on device geometries, investigate transmission profiles and I-V characteristics of test systems, and compare the numerical results with the reported experimental results. Finally, we summarize this paper in section \ref{sec:5}, emphasizing the simplicity and versatility of the multiscale modeling.

\section{\label{sec:theory}Theoretical framework}
\begin{figure}[ht!]
\begin{center}
\includegraphics[width=7.5cm,clip=true]{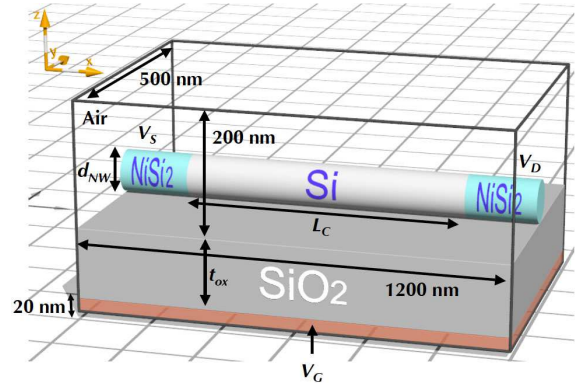}
\end{center}
\caption{\small{A model of a SiNW-based Schottky-barrier FET with realistic dimensions, as considered in this work. $V_\mathrm{S}$, $V_\mathrm{D}$, $V_\mathrm{G}$ are source, drain, and gate voltages, respectively. $L_\mathrm{c}$ is the length of the silicon nanowire  channel, and $t_\mathrm{ox}$ is the thickness of the oxide layer. $d_\mathrm{NW}$ is the diameter of the silicon nanowire. 
}}
\label{fig1}
\end{figure}

Figure \ref{fig1} shows a schematic image of a SiNW-based SB-FET. The device consists of a SiNW working as a channel and NiSi$_2$ segments working as source and drain contacts. This nanowire is put on the gate contact covered with SiO$_2$. The source, drain, and gate voltages, $V_\mathrm{S}$, $V_\mathrm{D}$, $V_\mathrm{G}$, the NW diameter $d_\mathrm{NW}$, as well as  the length of the silicon nanowire channel $L_\mathrm{c}$, and the thickness of the oxide layer $t_\mathrm{ox}$ are all indicated in figure \ref{fig1}. 

\begin{figure}[ht!]
\begin{center}
\includegraphics[width=8cm,clip=true]{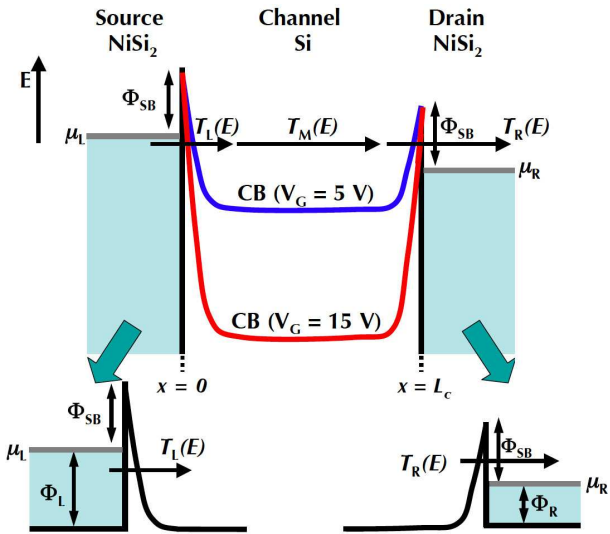}
\end{center}
\caption{\small{An energy diagram of a SiNW-based Schottky-barrier FET under an applied source-drain voltage with different gate voltages $V_\mathrm{G}$. The charge transport can be divided into 3 processes;  coherent tunneling through the Schottky-barrier $T_\mathrm{L/R}(E)$  at the left/right interfaces and ballistic transport $T_\mathrm{M}(E)$ in the middle. Increasing the gate voltage leads to stronger bending of the conduction band (CB), which reduces the width of the Schottky-barrier. $\Phi_{\mathrm{SB}}$ is the height of the Schottky-barrier. $\mu_{\mathrm{L/R}}$ and $\Phi_{\mathrm{L/R}}$ are the chemical potential and  the ground potential for left/right contacts, respectively.
}}
\label{fig2}
\end{figure}

The band diagram of the electronic transport process through a SB-FET under an applied source-drain voltage with different gate voltages is shown in figure \ref{fig2}. The energy gap $\Phi_{\mathrm{SB}}$ between the edges of the conduction band (CB) of the silicon channel and the Fermi energies of the NiSi$_2$ contacts is called the Schottky-barrier. The barrier thickness is reduced with  increasing gate voltage, allowing the electron in the source to tunnel through the potential energy barriers.\cite{Heinze,Avouris,Appenzeller} Thus, the drain current at the constant source-drain voltage is controlled by the gate voltage. 
The charge transport through the SB-FET can be divided into three stages (see  figure \ref{fig2}): 1. electron (hole) injection into the conduction band (valence band) by tunneling through the Schottky-barrier in the left interface; 2. ballistic or diffusive transport of charges through the SiNW; 3. tunneling of charges through the Schottky-barrier in the right interface.

According to figure \ref{fig2} we define the 1D potential energy profile $U(x)$ along the axis of the NiSi$_2$/Si/NiSi$_2$ nanowire as 


\[
  U(x)=  \begin{cases}
    \mu_{\mathrm{L}}-\Phi_{\mathrm{L}} & (x < 0) \\
    \mu_{\mathrm{L}}+\Phi_{\mathrm{SB}} +U_{\mathrm{ES}}  & (0 \le x \le L_{\mathrm{C}}) \\
    \mu_{\mathrm{L}} - eV_\mathrm {SD} - \Phi_{\mathrm{R}} & (L_\mathrm{C} < x)
  \end{cases}
\]
where the position $x = 0$ is set to be at the left NiSi$_2$/Si interface, $\mu_{\mathrm{L/R}}$ is the chemical potential of the left/right contact, $\Phi_{\mathrm{SB}}$ is the  Schottky-barrier energy, $\Phi_{\mathrm{L/R}}$ is the ground potential for left/right contact, $V_\mathrm {SD}$ is the applied source-drain voltage, and $U_\mathrm{ES}(x)$ is the 1D electrostatic potential along the axis of the Si-NW obtained from FEM calculations (see below). 
For the Schottky-barrier energy $\Phi_{\mathrm{SB}}$ we use experimental values from the literature (see below).
The ground potential  $\Phi_{\mathrm{L/R}}$ is introduced  in order to define the baseline for the transmission calculation at the two interfaces. The two ground potentials are chosen to span the bias window, i.e,  $\Phi_{\mathrm{L}}=eV_\mathrm{SD}^\mathrm{max}$, where $V_\mathrm{SD}^\mathrm{max}$ is the maximum source-drain (bias) voltage that is used in a set of calculations, and $\Phi_{\mathrm{R}}=0$.
This is how the energy of the electrons in the source that are within bias window is set higher than the ground potential. 
%
In order to calculate the charge transmission at the two interfaces (see bottom of figure \ref{fig2}) we use $U(x)$ in the interval $x=[-1.5 \mathrm{nm} \dots 21.5 \mathrm{nm}]$ for the left and $x=[L_\mathrm{c} - 21.5 \mathrm{ nm} \dots L_\mathrm{c} + 1.5 \mathrm{nm}]$ for the right interface. 
In these two ranges the potential inside the SiNW ($0 \le x  \le L_\mathrm{c}$) is regularized by applying an energy cutoff $U^\mathrm{cut}$ such that $U(x) = U^\mathrm{cut}$ for energies smaller than $U^\mathrm{cut}$. At the left (L) and right (R) interfaces the cutoff energy is  $U^\mathrm{cut}_\mathrm{L/R} = \mu_\mathrm{L/R} -\Phi_{\mathrm{L/R}}$, i.e., $U(x)$ is never smaller than the ground potential $\Phi_{\mathrm{L/R}}$ (see also figure \ref{fig5}).

In this study, we employed the Landauer approach in \textit{real-space} for the calculation of the electron transmission through the NiSi$_2$/Si interfaces.\cite{Meir,gdatta,bdatta,Appenzeller}
The Hamilton operator in 1D is given by  
\begin{equation*}
\hat{H}=-\frac{\hbar^2}{2m_\mathrm{eff}} \frac{\mathrm{d}^2}{\mathrm{d}x^2}+U(x),
\end{equation*}
where $m_{\mathrm{eff}}$, and $U(x)$ are the effective mass of electrons (or holes) and the  potential energy profile along the axis of the nanowire, respectively. 
Using  finite differences on an equally spaced grid, the Hamilton matrix is given by\cite{gdatta,bdatta,Datta2000}
\begin{equation*}
H_{n,m} = (U_n + 2t_0) \delta_{n,m} - t_0 \delta_{n,m+1} - t_0 \delta_{n,m-1},
\end{equation*}
where $U_n=U(x_n)$ is the potential, $x_n = an$ is the position, $a$ is the grid spacing, and $t_0=\hbar^2/2m_{\mathrm{eff}}a^2$.


The transmission functions at the left (L) and right (R) interfaces are obtained via the Landauer-B\"{u}ttiker formalism using the Fisher-Lee relation: 
\begin{equation*}
T_{\mathrm{L/R}}(E)= \mathrm{Tr}[ G^\mathrm{R} \Gamma_{\mathrm{L/R}}G^\mathrm{A}\Gamma_{\mathrm{M}}],
\end{equation*}
where Tr being the trace operation and $\Gamma_{\alpha}$ ($\alpha=\mathrm{L, R, M}$) are the broadening functions for the contacts, given by $\Gamma_{\alpha}(E) = i[ \Sigma_{\alpha}(E) - \Sigma_{\alpha}^\dag(E) ]$. $\Gamma_{\mathrm{M}}$ is the auxiliary broadening function, which is required because the system is cut into two 1D tunneling problems. 
The self-energies 
are defined as $\Sigma_{\alpha}(E)=-t_0\exp{(ika)}$, where $ka$ is obtained by inverting the band dispersion of the 1D wire (linear chain) $E(ka)=U_\alpha+2t_0 (1-\cos(ka))$, and $U_\alpha=U(x_\alpha)$ ($\alpha=\mathrm{L, R, M}$) is the potential at the position of the  left/right/auxiliary contacts.\cite{gdatta,bdatta,Datta2000} Note that the $\Sigma_{\alpha}$ is a diagonal matrix and its matrix elements are zero except for the position where the the left/right/auxiliary contact is attached. 

The retarded/advanced Green's functions $G^{\mathrm{R/A}}$ for the left (L) and right (R) interfaces are  defined as 
\begin{equation*}
G^{\mathrm{R/A}}(E)=[(E\pm i\eta)I - H-\Sigma_{\mathrm{L/R}} - \Sigma_{\mathrm{M}}]^{-1},
\end{equation*}
where  $i\eta$ is an infinitesimal imaginary value, $I$ is the identity matrix, $H=H_{n,m}$ is the Hamilton matrix, and $\Sigma_{\alpha}$ ($\alpha=\mathrm{L, R, M}$) are the self-energy matrices as defined above. 

Assuming ballistic charge transport through the silicon channel, i.e.\, $T_M(E)\simeq1.0$, and ignoring the phase memory, the total transmission through the device is given by\cite{gdatta} 
\begin{equation*}
T(E)=T_{\mathrm{L}}T_{\mathrm{R}}/(T_{\mathrm{L}}+T_{\mathrm{R}} - T_{\mathrm{L}}T_{\mathrm{R}}).
\end{equation*}
The assumption of ballistic transport in the core channel region was validated experimentally for devices having channel lengths less than 1 ${\mu}$m.\cite{weber}

Finally, the current through the SB-FET is calculated from 
\begin{equation*}
I_\mathrm{SD} = 2e/h\int_{-\infty}^{\infty}T(E)(F_{\mathrm{L}}(E)-F_{\mathrm{R}}(E)) \ dE.
\end{equation*} 
 The term $F_{\mathrm{L/R}}$ is the effective Fermi function $k$-summed over the transverse modes in the $y-z$ plane, represented by 
 \begin{equation*}
F_{\mathrm{L/R}}(E) = Sm_{\mathrm{eff}}k_{\mathrm{B}}T/\pi\hbar^2 \mathrm{ln}( 1+ \mathrm{exp} ( ( \mu_{\mathrm{L/R}} - E)/k_\mathrm{B}T)   ),
\end{equation*}
where $S$ and $\mu_{\mathrm{L/R}}$ are the cross-sectional area of SiNWs and the chemical potential for left/right electrodes, respectively. 
Under an applied source-drain voltage $V_\mathrm{SD}$, the chemical potential for the right electrode is given by $\mu_{\mathrm{R}}=\mu_{\mathrm{L}}-eV_\mathrm{SD}$. 

Note that the same formalism works for  hole currents by changing the band of interest from the conduction band (CB) to the valence band (VB), the Fermi function from $F_{\mathrm{L/R}}(E)$ to $1-F_{\mathrm{L/R}}(E)$, effective mass of electron to effective mass of hole, and exchanging the sign of the gate voltages.\cite{Jimenez07}

\begin{figure*}[ht!]
\begin{center}
\includegraphics[width=12cm,clip=true]{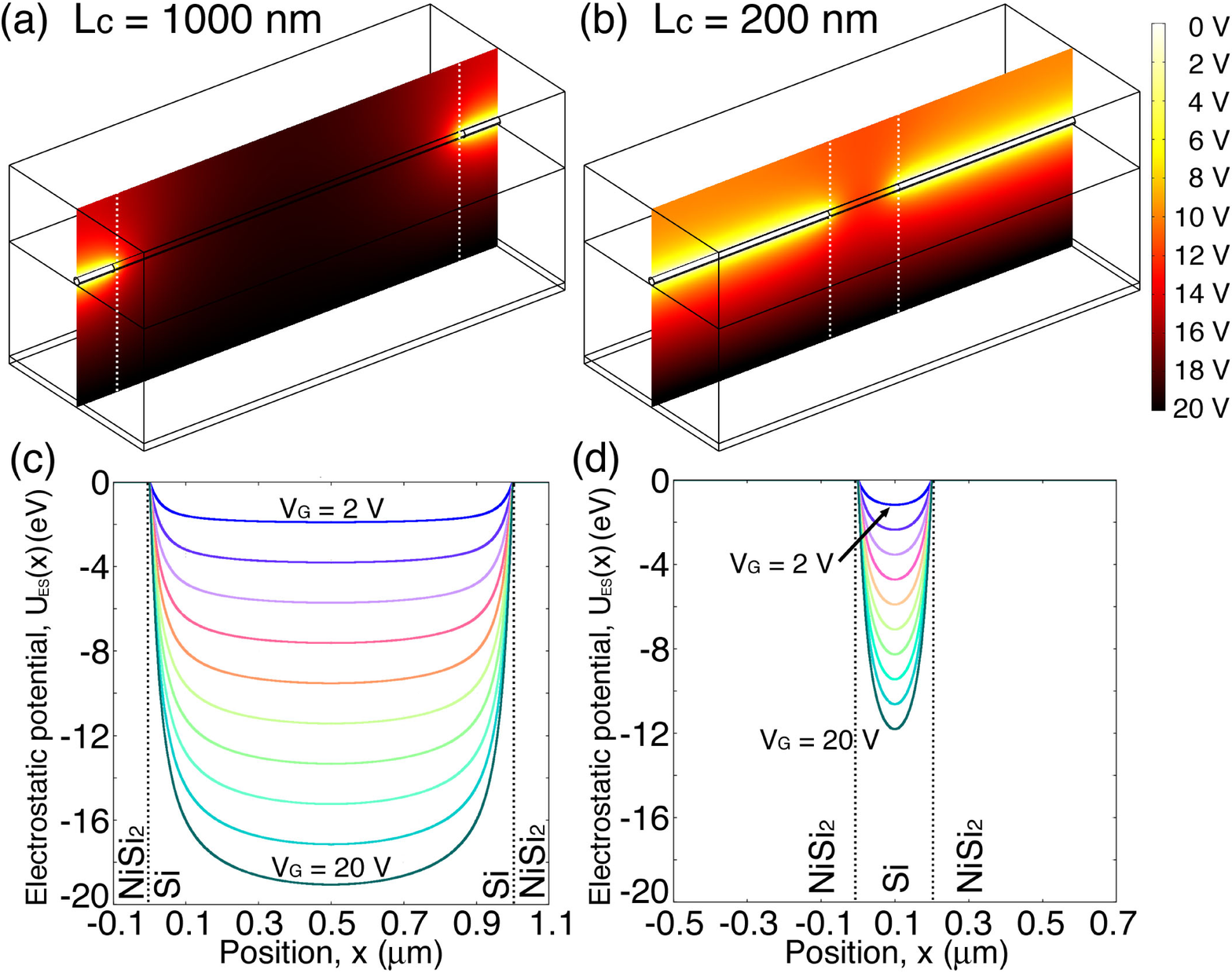}
\end{center}
\caption{\small{Electrostatic potential $U_\mathrm{ES}$ of the SiNW-based Schottky-barrier FETs with different channel lengths being $L_\mathrm{c} = 1000$ nm in (a)/(c) and $L_\mathrm{c} = 200$ nm in (b)/(d). 
Top: $U_\mathrm{ES}(x,z)$ in the $y=0$ plane; bottom: $U_\mathrm{ES}(x)$ along the axes of the corresponding wire. In these plots the metallic NiSi$_2$ nanowire is held at $U$ = 0 V.
The variation of the electrostatic potential is strongest close to the contacts.
}}
\label{fig3}
\end{figure*}

\section{\label{sec:methods}Computational details}
In this work we assumed that the effective electron mass of whole system for holes/electrons is equal to that of free electron. 
For all calculations we use $\mu_\mathrm{L}=0$ eV, $a$=1.0 \AA, $m_{\mathrm{eff}}=1.0 \times m_{\mathrm{e}}$=9.109$\times10^{-31}$ kg, and $T$= 300K. For the discussion of electron transport and figure \ref{fig3}-\ref{fig6} we use $t_\mathrm{ox}=300$ nm, $L_\mathrm{c}=1000$ nm, $d_\mathrm{NW}=20$ nm, $\Phi_{\mathrm{SB}} = 0.50$ eV, and $V^\mathrm{max}_\mathrm{SD}=0.5$ V, unless stated differently. For the discussion of hole transport and figure \ref{fig7}-\ref{fig8} we use $t_\mathrm{ox}=300$ nm, $d_\mathrm{NW}=21$ nm, $\Phi_{\mathrm{SB}} = 0.44$ eV, $V^\mathrm{max}_\mathrm{SD}=2$ V.
The 3D electrostatic potential $U_\mathrm{ES}(x,y,z)$ for each combination of  $V_{\mathrm{L}}$, $V_{\mathrm{R}}$, and $V_{\mathrm{G}}$ is calculated by solving Poisson equation with a commercial FEM software (COMSOL multiphysics.\cite{comsol}) The boundary potentials for the source, drain and gate contacts are kept constant at $V_{\mathrm{L}}$, $V_{\mathrm{R}}$, and $V_{\mathrm{G}}$, respectively.  The finite element mesh for the modeled structures is automatically generated  with controlled distribution and increased density in the regions close to the insulator and NiSi$_2$/Si interfaces. The 1D potential profile $U_\mathrm{ES}(x)$ along the axis of the NW is extracted from $U_\mathrm{ES}(x,y,z)$.

\section{\label{sec:4}Result and discussion}


\subsection{Dependence of the electrostatic potential on device geometries}

As a first examination of the relationship between the geometry of FET devices and the efficiency of the gate effect, we have calculated 3D electrostatic potentials of FET devices. 
Figure \ref{fig3}(a) and (b) depict the potential landscapes 
with a gate potential of $V_\mathrm{G} = 20$ V and the absence of a source-drain voltage.  Figure \ref{fig3}(c) and (d) show the same potential along the axes of the channels for long and short silicon channels, respectively. In figure \ref{fig3}(a) and (c), the potential along the silicon channel drops strongly due to the applied gate field, whereas the gate field does not penetrate efficiently in the shorter silicon channel in figure \ref{fig3}(b) and (d).

\begin{figure}[ht!]
\begin{center}
\includegraphics[width=8cm,clip=true]{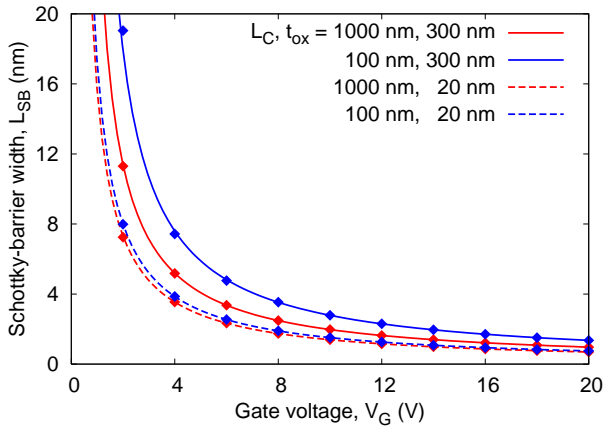}
\end{center}
\caption{\small{Dependence of Schottky-barrier width $L_\mathrm{SB}$ on the gate voltage $V_\mathrm{G}$, the channel length $L_\mathrm{c}$, and the thickness of the oxide layer $t_{\mathrm{ox}}$. $L_\mathrm{SB}$ is defined in figure \ref{eqn:SBW}. The continuous lines are fits of the data points to $L_{\mathrm{SB}}=c V_{\mathrm{G}}^{-1}$, i.e., the Schottky-barrier width is inversely proportional to the gate voltage. The gate effect is most efficient for large channel lengths and thin oxide layers.
}}
\label{fig4}
\end{figure}

In addition we have analyzed the influence of the thickness of the insulating layer between the nanowire device and the gate on the Schottky-barrier width. 
The Schottky-barrier width $L_{\mathrm{SB}}$ of the left interface  is defined by the position $x=L_{\mathrm{SB}}>0$ satisfying  
\begin{equation}
U(L_{\mathrm{SB}})=\mu_\mathrm{L}.
\label{eqn:SBW}
\end{equation} 
In figure \ref{fig4}, we show the Schottky-barrier width of the FET devices 
as a function of applied gate voltage with varying thickness of the gate insulators and the channel lengths.
We can see that the gate field reduces the Schottky-barrier width for all devices. We find that the Schottky-barrier width is inversely proportional to the gate voltage. The reduction of the silicon channel length as well as an increase of the width of the oxide layer prevents the gate field from efficiently penetrating into the Schottky-contact, leaving the Schottky-barrier width thicker than the device with long channel or an thick oxide layer. 
%
Since the gate voltage narrows the Schottky barrier the enhancement of the electron tunneling by the gate field is larger through SB-FET devices with longer silicon channels and thinner gate insulators.

\subsection{Transmission function and I-V characteristics}

\begin{figure*}[ht!]
\begin{center}
\includegraphics[width=13cm,clip=true]{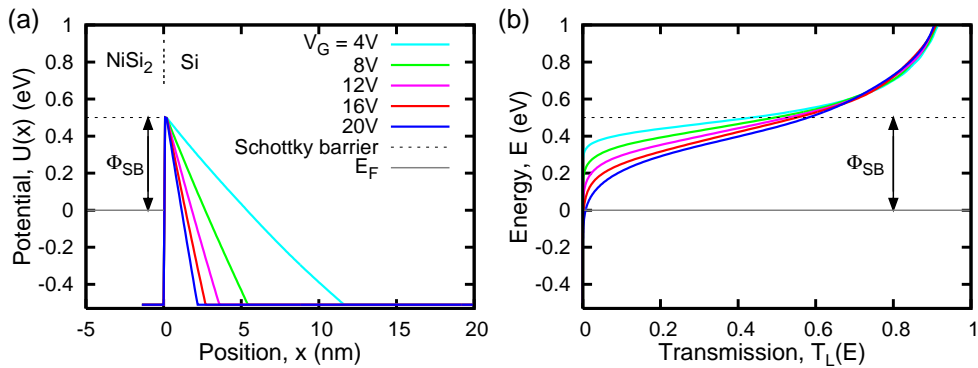}
\end{center}
\caption{\small{(a) The potential profile $U(x)$ across the central axis of the left NiSi$_2$/Si interface for different gate voltages $V_\mathrm{G}$, and (b) the corresponding transmission functions $T_\mathrm{L}(E)$. Here the Schottky-barrier height  $\Phi_{\mathrm{SB}} = 0.50$ eV.
}}
\label{fig5}
\end{figure*}

Figure \ref{fig5} shows potential profiles $U(x)$ at the left interface for different gate voltages and the corresponding transmission functions $T_\mathrm{L}(E)$ for \textit{electron} transmissions through these Schottky-barriers. The non-zero transmissions are due to the quantum tunneling effect. Since a positive gate voltage causes the Schottky-barrier to get thinner, the  electron transmission increases with increasing gate voltage. For electrons with energies above the Schottky-barrier height $\Phi_{\mathrm{SB}}$, the electron transmission quickly approaches its maximum value 1. Therefore our model takes into account both tunneling currents and thermo-activated (thermionic) currents.
  
\begin{figure*}[ht!]
\begin{center}
\includegraphics[width=13cm,clip=true]{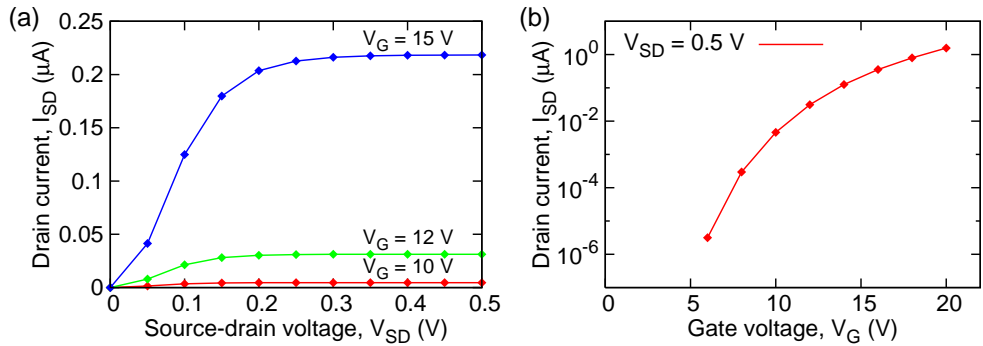}
\end{center}
\caption{\small{ Electron transport characteristics of the SB-FET device. (a) Drain current $I_\mathrm{SD}$ versus source-drain voltage $V_\mathrm{SD}$ for different gate voltages $V_\mathrm{G}$. The drain current saturates at higher bias voltages. The saturated current is significantly enhanced with  higher gate voltages. (b) Drain current $I_\mathrm{SD}$ versus gate voltage $V_\mathrm{G}$ for a fixed source-drain voltage ($V_\mathrm{SD}$ = 0.5 V). We obtain the typical behavior of a conventional FET device. 
}}
\label{fig6}
\end{figure*}

After calculating the transmission functions for both left and right interfaces, we have calculated the I-V characteristics of SB-FET devices by integrating over the total transmission function. 
Figure \ref{fig6}(a) shows the calculated drain current $I_\mathrm{SD}$ versus source-drain voltage $V_\mathrm{SD}$ for different gate voltages. The I-V curves show typical features of conventional FETs, i.e., a linear increase of current followed by current saturation for higher source-drain voltages. The saturation currents are strongly enhanced with increasing gate fields leading to a typical switching behavior with a high on/off-current ratio. We evaluated this behavior by calculating the drain current at a fixed source-drain voltage for several gate voltages. Figure \ref{fig6}(b) presents the saturated drain current versus the gate voltage for a fixed source-drain voltage $V_{\mathrm{SD}}$ = 0.5 V. 
The drain current increases exponentially with the increase of the gate field  in the lower gate voltages. This is due to the rapid reduction of the Schottky-barrier width with the increase of gate fields as shown in the figure \ref{fig4}. The saturation of the drain current in higher gate field can also be explained from the figure \ref{fig4} since the Schottky-barrier width does not decrease so much in the higher gate field. 
The huge on/off-current ratio implies that the device is promising for logic operations. The flatness of the current in the saturation region in figure \ref{fig6}(a) is also beneficial for the logic operation since the current retains constant with respect to fluctuation of the source-drain voltage.

\subsection{Comparison with experiment}

Many measurements of SB-FETs consisting of SiNWs have exhibited unipolar p-type transfer characteristics,\cite{weber,weber2,weber3,weber4,Appenzeller} thus the current is transported by holes in these systems. As discussed previously, 
our model can also be applied to hole transport systems. In order to check the validity of the model, we have applied our model to SiNW-based SB-FETs and compared the I-V characteristics and on/off-current ratios with the experimental results measured by  Weber \textit{et~al.~}\cite{weber} 

The numerical parameters are set as follows. The Schottky-barrier height between Ni-silicide and bulk Si ranges from $0.35$ eV to $0.47$ eV for holes.
\cite{Rees_and_Matthai} Thus, we set the Schottky-barrier height of the NiSi$_2$/Si interface for holes as $\Phi_{\mathrm{SB}}$= 0.44 eV. The diameter of the SiNWs is set to $d_{\mathrm{NW}}$ = 21 nm, identical to the mean diameter of the literature.\cite{weber} The oxide thickness on the gate is set to $t_{\mathrm{ox}} = $300 nm. 

\begin{figure}[ht!]
\begin{center}
\includegraphics[width=7.5cm,clip=true]{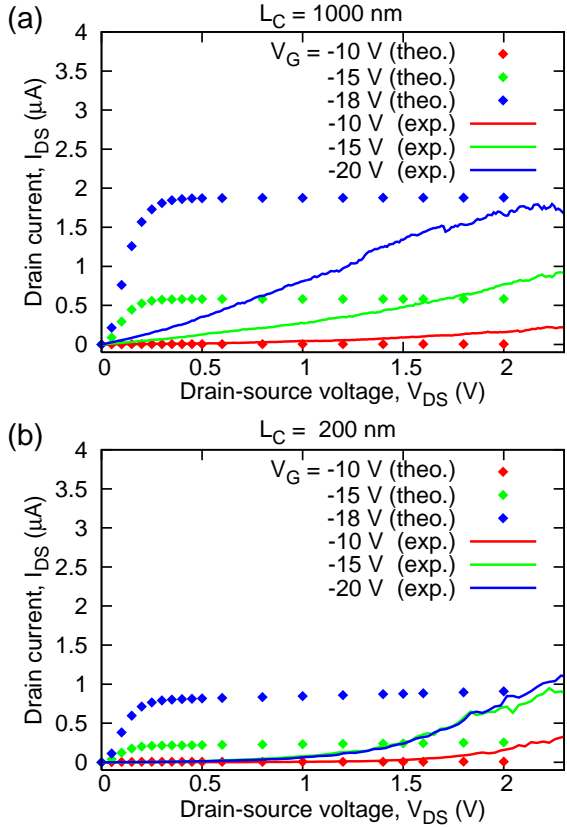}
\end{center}
\caption{\small{I-V characteristics of two FET devices for (a) a long and (b) a short channel length at different gate voltages. The calculated currents are shown with dots. The experimental results given by Weber \textit{et al.}\cite{weber} are shown as continuous lines for comparison. Although the theoretical currents reach their saturations at lower bias, the saturated currents show a good agreement with experiments. Note that $I_{\mathrm{DS}}=-I_{\mathrm{SD}}$ and $V_{\mathrm{DS}}=-V_{\mathrm{SD}}$.
}}
\label{fig7}
\end{figure}

Figure \ref{fig7} presents the calculated drain currents $I_{\mathrm{SD}}$ versus source-drain voltage $V_{\mathrm{SD}}$ for (a) long ($L_{\mathrm{c}}=$ 1000 nm) and  (b) short gate length ($L_{\mathrm{c}}=$ 200 nm) with experimental data provided by Weber \textit{et al.}\cite{weber} The calculated I-V curves for both systems exhibit similar features as the experimental results having both a linear increase and  saturation at higher source-drain voltages.
The drain current through the shorter channel is smaller than the one through the longer channel since the gate field does not optimally penetrate the NiSi$_2$/Si interfaces.
Although the slopes of the calculated I-V curves for low source-drain voltages significantly exceed experimental ones, the amount of the saturated drain current shows a good agreement with the experiment except for the currents with high gate field.

\begin{figure}[h]
\begin{center}
\includegraphics[width=7.5cm,clip=true]{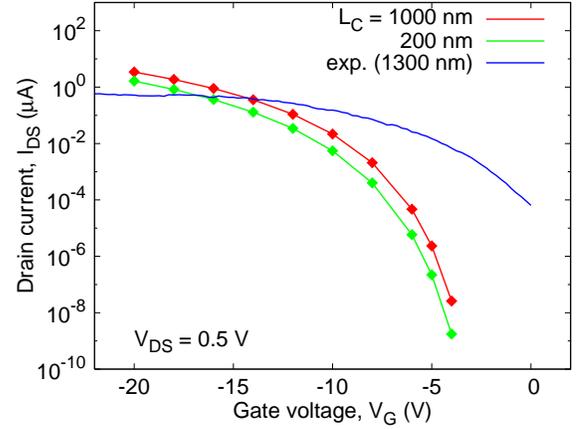}
\end{center}
\caption{\small{Drain current $I_\mathrm{DS}$ versus gate voltage $V_\mathrm{G}$ characteristics of SiNW-based SB-FET devices with long (in red) and short (in green) silicon channels. The source-drain voltage is fixed to be $V_\mathrm{DS}$ = 0.5 V. For comparison, the experimental results of Weber \textit{et al.~}\cite{weber} for a  long (1300 nm) SiNW are shown (in blue). Numerical results show a qualitative agreement with experiment. The drain currents of the SB-FET with the long channel are higher than for the short channel because the gate effect is more efficient.
}}
\label{fig8}
\end{figure}

Figure \ref{fig8} presents the saturated drain current versus the gate voltage of SB-FET devices with long and short channels. The saturated drain currents increase significantly with the increase of the gate field since the gate field reduces Schottky-barrier width as shown figure \ref{fig4}. The on-currents in the short and long systems are of the same order of magnitude. On the other hand, the off-currents for small gate voltages are significantly smaller for the short channel. 
Although the drain current in the lower gate field is underestimated, the numerical results show a qualitative agreement with experiment. 
Generally we observe that the drain currents of the SB-FET with the long channel are higher than for the short channel because the gate effect is more efficient (compare figure \ref{fig3}(c), (d), figure \ref{fig4}, and figure \ref{fig7}(a), (b)).

The discrepancies between our model and the experiment come from the simplicity of our model.  First, we did not take into account the sub-bands\cite{JWang,Dubois,Kosina} but  expressed the transport by using a single band model. The inclusion of the sub-bands  will lower the drain current for small source-drain voltages since the onset of current flow through the channel will be shifted to higher source-drain voltages due to the difference of energies between sub-bands. 
In addition, the reconstruction of the exact potential profile at the Schottky-contact\cite{Andreas,Stiles} due to space-charge effects and the dependence of the potential on the lateral position in the wire are not considered in this study. Furthermore, the effect of capacitance at the left and right interfaces and at the gate contact are not included.\cite{Sajjad} The addition of a native SiO$_2$ layer on the surface of the nanowire and the corresponding change of the electronic structures between the core and surface of the nanowire will also have an impact on the transport properties.\cite{Rurali-2011} In order to capture the underlying physics of SB-FET devices, an advanced model incorporating these effects will be needed. Further studies including these effects will give helpful information for the future development of nanowire-based FET devices.


\section{\label{sec:5}Conclusion}


In summary, 
nanowire-based FETs have been focusing attention as a promising platform for sensor applications due to their high sensitivity. In order to understand the origin of the physical behavior in the nanowire-based FET devices, we have developed a multiscale model combining classical FEM and the Landauer approach.
Our model is conceptually simple, computationally inexpensive, it uses only a few empirical parameters, and it is possible to simulate devices with dimensions ranging from a few nanometers up to some micrometers.
We have applied this model to Schottky-barrier FETs consisting of SiNWs working as channel and NiSi$_2$ NW working as source and drain contacts. 
Our calculated I-V characteristics  showed the typical behavior of conventional FET devices, having a linear increase of current followed by a saturation. The saturated drain currents increase with increasing gate field due to the reduction of the Schottky-barrier width. Despite the simplicity of the model, the calculations showed a good agreement with experiments and the model correctly reproduces the dependence of the saturated drain current on the gate field and the device geometry.

Our approach is suitable for one-dimensional Schottky-barrier field-effect transistors of arbitrary device geometry. The model can be improved by taking into account for example insulating layers surrounding the nanowires, the deformation of the electrostatic potential at the Schottky-contacts, or accurate electronic band dispersions. The results obtained from our model will serve as guidelines for the understanding of ongoing experimental work and will help in the design of device architectures for future applications.

%
%

\section*{Acknowledgments}
This work was supported by the European Union (European Social Fund)
 and the Free State of Saxony (S\"achsische Aufbaubank) in the young
researcher group 'InnovaSens' (SAB-Nr. 080942409). This work was also supported by the Volkswagen Foundation and by the WCU (World  Class University) program through the Korea Science and Engineering Foundation funded by the Ministry of Education, Science and Technology (Project No. R31-2008-000-10100-0). We acknowledge the Center for Information Services and High Performance Computing (ZIH) at the Dresden University of Technology for  computational resources. We thank  Walter Weber (NamLab), Cormac Toher (TU Dresden) and Stephan Roche (CEA, INAC, and CIN2(ICN-CSIC)) for useful discussions. 

\section*{References}

\begin{thebibliography}{10}

\bibitem{Wagner}
Wagner R S and Ellis W C
\newblock {\em Appl. Phys. Lett.} 1964 \textbf{4} 89 

\bibitem{Westwater}
J.~Westwater, D.~P. Gosain, S.~Tomiya, S.~Usui, and H.~Ruda.
\newblock {\em J. Vac Sci. Technol. B} 1997 \textbf{15} 554

\bibitem{CP2010}
V.~Schmidt, J.~V. Wittemann, and U.~Goesele.
\newblock {\em Chem. Rev.} 2010 \textbf{110}  361

\bibitem{Rurali}
Rurali R
\newblock {\em Rev. Mod. Phys.} 2010 \textbf{82} 427

\bibitem{Cui}
Cui Y, Wei Q Q, Park H K and Lieber C M
\newblock {\em Science} 2001 \textbf{293} 1289

\bibitem{APA}
Li Z, Rajendran B, Kamins T I, Li X, Chen Y and  Williams R S
\newblock {\em Appl. Phys. A} 2005 \textbf{80} 1257

\bibitem{photo}
Servati P, Colli A, Hofmann S, Fu Y Q, Beecher P,  Durrani Z A K, 
  Ferrari A C, Flewitt A J, Robertoson J and Milne W I
\newblock {\em Physica E} 2007 \textbf{38} 64

\bibitem{bipo}
Cui Y and Lieber C M
\newblock {\em Science} 2001 €textbf{291} 851 

\bibitem{Koo}
Koo S M, Edelstein M D, Li Q, Richter C A  and Vogel E M
\newblock {\em Nanotechnology} 2005 \textbf{16} 1482

\bibitem{weber}
Weber W M, Geelhaar L, Graham A P, Unger E, Duesberg G S, Liebau M, Palmer W, Cheze C, Riechert H, Lugli P and Kreupl F
\newblock {\em Nano. Lett.} 2006 \textbf{6} 2660

\bibitem{weber2}
Weber W M, Graham A P, Duesberg G S, Liebau M, Cheze C, Geelhaar L,Unger E, Pamler W, Hoenlein W, Riechert H, Kreupl F and Lugli P
\newblock {\em ESSDERC 2006. Proceeding of the 36th European} 2006 423

\bibitem{weber3}
Weber W M, Geelhaar L, Unger E,  Cheze C, Kreupl F, Riechert H and Lugli P
\newblock {\em Phys. Stat. Sol. (b)} 2007 \textbf{244} 4170

\bibitem{weber4}
Weber W M, Geelhaar L, Lamagna L, Fanciulli M, Kreupl F, Unger E, Riechert H, Scarpa G and Lugli P
\newblock {\em Nanotechnology, 2008. NANO '08. 8th IEEE Conference} 2008 580

\bibitem{Nano_Lett-04-51}
Hahm J-I and Lieber C M
\newblock {\em Nano Lett.} 2004 \textbf{4} 41

\bibitem{Nano_Lett-04-245}
Li Z, Chen Y, Li X, Kamins T I, Nauka K and Williams R S
\newblock {\em Nano Lett.} 2004 \textbf{4} 245

\bibitem{Nano_Lett-08-1066}
 Zhang G-J, Zhang G, Chua J H, Chee R-E, Wong E H, Agarwal A, Buddharaju K D, Singh N, Gao Z and Balasubramanian N
\newblock {\em Nano Lett.} 2008  \textbf{8} 1066

\bibitem{Anal_Chem-79-3291}
Gao Z, Agarwal A,  Trigg A D, Singh N, Fang C, Tung C-H, Fan Y, Buddharaju K D and Kong J
\newblock {\em Anal. Chem.} 2007 \textbf{79} 3291

\bibitem{electrochem}
Scheibal Z R, Xu W, Audiffred J F, Henry J E and  Flake J C.
\newblock {\em Electrochem. Solid-State Lett.} 2008 \textbf{11} K81

\bibitem{Biosens_Bioelec-accepted}
Chen C C, Chen Y-Z, Huang Y-J and Sheu J-T
\newblock {\em Biosens. Bioelectron.} 2011 \textbf{26} 2323

\bibitem{Knoch1}
Knoch J, Lengeler B and Appenzeller J
\newblock {\em IEEE Trans. Elec. Dev.} 2002 \textbf{49} 1212

\bibitem{Knoch2}
Knoch J, Lengeler B and Appenzeller J
\newblock {\em Appl. Phys. Lett.} 2002 \textbf{81} 3082

\bibitem{Heinze}
Heinze S, Tersoff J, Martel R, Derycke V, Appenzeller J and Avouris Ph
\newblock {\em Phys. Rev. Lett.} 2002 \textbf{89} 106801

\bibitem{Pourfath}
Pourfath M, Ungersboeck E, Gehring A, Kosina H, Selberherr S,  Park W J
  and Cheong B H
\newblock {\em J. Comp. Electron} 2005 \textbf{4} 75

\bibitem{Michetti}
Michetti P and Iannaccone G
\newblock {\em IEEE-NANO 9th IEEE Conference} 2009 25

\bibitem{Appenzeller}
Appenzeller J, Knoch J, Tutuc E, Reuter M and Guha S.
\newblock {\em IEDM Tech. Dig.} 2006 1

\bibitem{Jimenez07}
Jimenez D, Cartoixa X, Miranda E, Sune J, Chaves F A and Roche S
\newblock {\em Nanotechnology} 2007 \textbf{18} 025201 

\bibitem{Dubois}
Dubois M, Jimenez D, de~Andres P L  and Roche S
\newblock {\em Phys. Rev. B} 2007 \textbf{76} 115337

\bibitem{Landauer}
Landauer R
\newblock {\em IBM J. Res. Dev.} 1957 \textbf{1} 233

\bibitem{Keldysh}
Keldysh L V
\newblock {\em Zh. Eksp. Teor. Fiz.} 1964 \textbf{47} 1515

\bibitem{Buttiker}
B\"{u}ttiker M
\newblock {\em Phys. Rev. Lett.}, \textbf{57} 1761 

\bibitem{Avouris}
Avouris Ph
\newblock {\em Acc. Chem. Res.} 2002 \textbf{35} 1026

\bibitem{Meir}
Meir Y and Wingreen N S
\newblock {\em Phys. Rev. Lett.} 1992 \textbf{68} 2512

\bibitem{gdatta}
Datta S 1995
\newblock {\em Electronic Transport in Mesoscopic Systems}
\newblock (Cambridge: Cambridge University Press, Cambridge)

\bibitem{bdatta}
Datta S 2005
\newblock {\em Quantum Transport: Atom to Transistor}
\newblock (Cambridge: Cambridge University Press)

\bibitem{Datta2000}
Datta S
\newblock {\em Superlattice Microst.} 2000 \textbf{28} 253

\bibitem{comsol}
http://www.comsol.com

\bibitem{Rees_and_Matthai}
Rees N V and Matthai C C
\newblock {\em Semicond. Sci. Tech.} 1989 \textbf{4} 412

\bibitem{JWang}
Wang J, Rahman A, Ghosh A, Klimeck G and Lundstrom M
\newblock {\em Appl. Phys. Lett.} 2005 \textbf{86} 093113

\bibitem{Kosina}
Neophytou N and Kosina H
\newblock {\em Nano Lett.}  \textbf{10} 4913

\bibitem{Andreas}
de~Andres P L, Garcia-Vidal F J, Reuter K and Flores F
\newblock {\em Prog. Surf. Sci.} 2001 \textbf{66} 3

\bibitem{Stiles}
Stiles M D and Hamann D R
\newblock {\em Phys. Rev. B.} 1989 \textbf{40} 1349

\bibitem{Sajjad}
Sajjad R N, Alam K and Khosru Q D M
\newblock {\em Semicond. Sci. Tech.} 2009 \textbf{24} 045023

\bibitem{Rurali-2011}
Amato M, Ossicini S and Rurali R
\newblock 2011 \textit{Nano Lett.} \textbf{11} 594

\end{thebibliography}

\end{document}